\documentclass[fleqn,usenatbib]{mnras}

\usepackage[utf8]{inputenc}
\usepackage{threeparttable}
\usepackage{amsmath}
\usepackage{epstopdf}
\usepackage{threeparttable}
\usepackage{graphicx, color, url}
\usepackage{bm}
\usepackage{soul}
\usepackage{amssymb} 
\usepackage{booktabs}
\usepackage[usenames,dvipsnames,svgnames,table]{xcolor}
\definecolor{darkblue}{rgb}{0.0,0.0,0.3}
\hypersetup{colorlinks,%
            linkcolor=darkblue,urlcolor=darkblue,
            anchorcolor=darkblue,citecolor=blue} 
\usepackage{txfonts}

\newcommand{\eqb}{\begin{eqnarray}}
\newcommand{\eqe}{\end{eqnarray}}

\defcitealias{christie2019}{C19}
\defcitealias{pgs16}{PGS16}

\title[Inter-Plasmoid Compton Scattering]{Inter-Plasmoid Compton Scattering and the Compton Dominance of BL Lacs}

\author[]{I.M.~Christie$^1$\thanks{E-mail: ichristi231@gmail.com}, M.~Petropoulou$^2$\thanks{E-mail: m.petropoulou@astro.princeton.edu}, L.~Sironi$^{3}$, \& D.~Giannios$^{4,5,6}$ \\
$^1$Center for Interdisciplinary Exploration \& Research in Astrophysics (CIERA), Physics \& Astronomy, Northwestern \\ University, Evanston, IL 60208, USA \\
$^2$Department of Astrophysical Sciences, Princeton University, 4 Ivy Lane, Princeton, NJ 08544, USA \\
$^3$Department of Astronomy and Columbia Astrophysics Laboratory, Columbia University, 550 W 120th Street, New York, NY 10027, USA \\
$^4$Department of Physics, Purdue University, 525 Northwestern Avenue, West Lafayette, IN, 47907, USA\\
$^5$Department of Physics, University of Crete, Voutes, GR-70013, Heraklion, Greece\\
$^6$Institute of Astrophysics, Foundation for Research and Technology Hellas, Voutes, GR-70013, Heraklion, Greece}

\begin{document}
\setstcolor{red}

\date{Received.../Accepted...}

\pagerange{\pageref{firstpage}--\pageref{lastpage}} \pubyear{2019}

\maketitle

\label{firstpage}

\begin{abstract}
Blazar emission models based on magnetic reconnection succeed in reproducing many observed spectral and temporal features, including the short-duration luminous flaring events. 
Plasmoids, a self-consistent by-product of the tearing instability in the reconnection layer, can be the main source of blazar emission. 
Kinetic simulations of relativistic reconnection have demonstrated that plasmoids are characterized by rough energy equipartition between their radiating particles and magnetic fields. This is the main reason behind the apparent shortcoming of plasmoid-dominated emission models to explain the observed Compton ratios of BL Lac objects.
Here, we demonstrate that the radiative interactions among plasmoids, which have been neglected so far, can assist in alleviating this contradiction.
We show that photons emitted by large, slow-moving plasmoids can be a potentially important source of soft photons to be then up-scattered, via inverse Compton, by small fast-moving, neighboring plasmoids.
This inter-plasmoid Compton scattering process can naturally occur throughout the reconnection layer, imprinting itself as an increase in the observed Compton ratios from those short and luminous plasmoid-powered flares within BL Lac sources, while maintaining energy equipartition between radiating particles and magnetic fields.
\end{abstract}


\section{Introduction}
\label{sec:intro}

Blazars constitute a small subclass of active galactic nuclei (AGN) in which their relativistic jets are pointed towards our line of sight \citep{Blandford1978}. They are well-known for the double-humped appearance of their spectral energy distribution \citep[SED;][]{Urry1995}, their multi-timescale (i.e. from minutes to months) and multi-wavelength variability  \citep[e.g.][]{Maraschi_1999,Fossati_2008, Albert2007,Aharonian2007,Ackermann2015,Ahnen2016, Ahnen_2017}. BL Lac sources, which are traditionally distinguished from other AGN subclasses by their weak or even absent  optical emission lines, in particular, are observed to have Compton ratios (i.e. luminosity ratio of the high-to-low energy components, hereafter denoted as $A_C$) in the range of $\sim 0.2 - 2$  \citep{Ghisellini2010,fincke2013, Nalewajko2017,Padovani2017}.

\begin{figure*}
\centering
\hspace{1cm}
\includegraphics[height=0.55\textwidth]{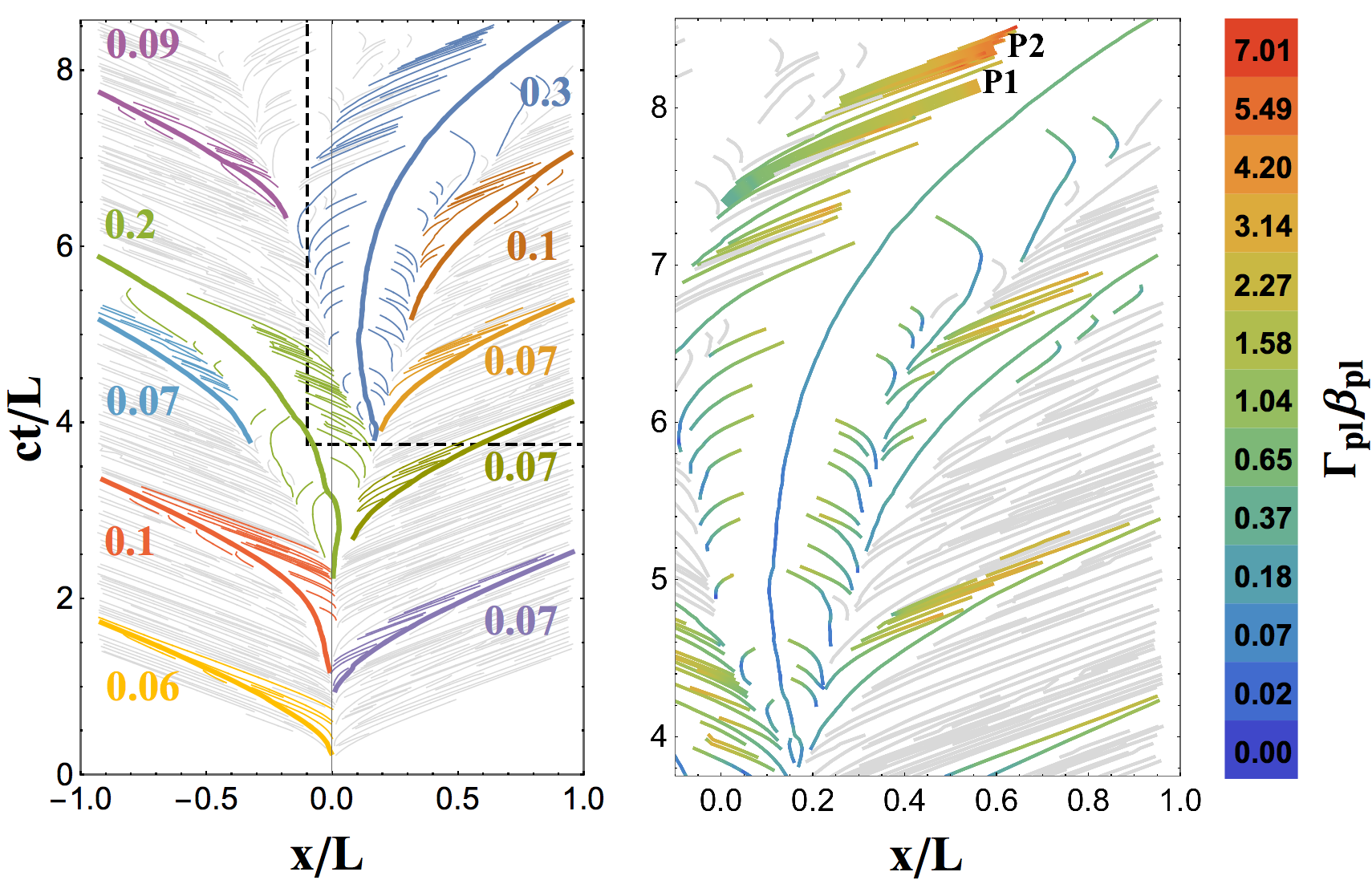}
\caption{Position-time diagram of the reconnection layer adopted from a $\sigma = 10$ PIC simulation of reconnection in pair plasma \citep{sgp16}. 
Each track shows the location of a plasmoid center (in units of the layer's half-length $L$) at different times (in units of $L/c$) measured in the layer's rest frame. The left panel highlights several subsets of plasmoids in which smaller plasmoids (sizes $\sim 0.008-0.05 \, L$) trail behind, and eventually coalesce with, a larger one  denoted by a thick line with the same color (its final size, in units of $L$, is also marked on the plot). The right panel displays a zoomed-in portion of  the layer (box-shaped region in the left panel) with the color denoting the plasmoid dimensionless four-velocity $\Gamma_{\rm pl} \, \beta_{\rm pl}$. Two representative plasmoids, whose photon emission is computed with inclusion of IPCS (see Fig.~\ref{fig:small_plasmoids_SED_LC}), are denoted as P1 and P2. 
A coloured version of this plot is available online.}
\label{fig:sigma_10_tracks}
\end{figure*}

Spherical blobs of plasma containing magnetic fields and relativistic non-thermal particles have been invoked to explain the spectral and temporal features observed from many blazars in flaring and non-flaring states \citep{Dermer_1992, Sikora_1994, Mastichiadis1995,bloom1996,Chiaberge1999,Celotti2008,Ghisellini2010}. This modeling offers a probe of the physical conditions in the emitting regions and, in particular, of the relative energy densities of radiating particles and magnetic fields.
For one-zone emission models (i.e. in which only a single region is responsible for the production of the observed multi-frequency emission) of BL Lac sources, studies have shown that the energy density of the electrons in the blazar zone typically exceeds that of the magnetic field by a factor of $\gtrsim 10-10^3$ \citep{Tavecchio2010, Tavecchio_2016}. To combat this discrepancy, anisotropies in the relativistic non-thermal particle distribution within the emitting regions have also been invoked \citep{Sobacchi2019,Tavecchio2019}.

This apparent dominance of particles over magnetic fields is at odds with the jet composition at its base.
Relativistic jets in AGN are believed to be launched as Poynting-flux dominated flows \citep{Blandford1977,Tchekhovskoy2009,Tchekhovskoy2011} with magnetization\footnote{Plasma magnetization is the ratio of magnetic to particle energy density: for a cold pre-reconnection plasma, $\sigma = B^2 /(4 \pi \rho c^2)$, where $B$ and $\rho$ are the magnetic field and particle mass density.} $\sigma \gg 1$. The presence of emitting regions within the magnetically dominated relativistic blazar jet has since motivated numerical and analytical studies of magnetic reconnection \citep[e.g.][]{Spruit2001,giannios2006,giannios2009,Nalewajko2011,Nalewajko2018}, a process which imparts a fraction of the jet's magnetic energy to energetic particles. In the relativistic regime of reconnection (i.e. $\sigma > 1$), most applicable to blazar jets, quasi-spherical magnetized plasma structures containing magnetic fields and non-thermal particles -- the so-called \textit{plasmoids} -- naturally form and move relativistically within the reconnection layer \citep[e.g.][]{Uzdensky2010,sironi2014,guo2014,guo2015,sgp16,Werner2016,Werner2018}. The reconnection-generated plasmoids are promising physical candidates for the blobs invoked to power the blazar emission \citep[][hereafter, the latter two will be denoted as \citetalias{pgs16} and \citetalias{christie2019}, respectively]{giannios2013, pgs16,christie2019}.
In the absence of a strong non-reconnecting magnetic field component (the so-called ``guide field''), relativistic particles and magnetic fields within plasmoids are in rough energy equipartition \citep[e.g.][]{spg15, Werner2018, petropoulou_2019}, even though the magnetic field largely dominates (i.e. $\sigma>1$) in the unreconnected plasma.

The dynamics and properties of plasmoids formed during the reconnection process can be studied from first principles with particle-in-cell (PIC) simulations. 
We have recently incorporated the results of two-dimensional (2D) PIC simulations of reconnection into radiative models of blazar emission (\citetalias{pgs16}, \citetalias{christie2019}). One of the major findings of these studies is that numerous plasmoids forming during a single reconnection event can naturally produce multi-wavelength and multi-timescale variability, 
including extreme flares of short durations \citep[$\sim$min timescales, see e.g.][]{Aharonian2007}. In addition, the cumulative emission of several plasmoids, at any given time, exhibits the usual two-hump structure observed in blazar SEDs.

Despite the success in reproducing the global features of blazar emission, plasmoid-based emission models for BL Lac sources result in $A_C \lesssim 0.1$, namely they have difficulty in reproducing the observed Compton ratios (see Figs.~10 and~12 in \citetalias{pgs16}, Figs.~6 and~8 in \citetalias{christie2019}, and \citet{Morris2019}). This apparent shortcoming of the models comes mainly from the combination of two factors: ({\it i}) the adoption of equipartition between relativistic particles and magnetic fields within plasmoids, as informed by PIC simulations of reconnection; and ({\it ii}) the assumption that no  photon fields are available for inverse Compton up-scattering other than the synchrotron photons produced by the plasmoid itself (photon sources external to the jet are negligible in BL Lac objects). 
If, however, other photon sources were to be available for up-scattering, then it would be possible to obtain larger Compton ratios without invoking deviations from equipartition.
 
In this work, we study for the first time the radiative interactions among plasmoids within the same reconnection layer.
In particular, we demonstrate that most small and mid-sized plasmoids trail behind and eventually coalesce with a larger slow-moving plasmoid. Their relative motions can be relativistic, thus making the Doppler-boosted synchrotron radiation of large plasmoids a potential source of soft photons for Compton scattering by the pairs in  smaller trailing plasmoids  (see also discussion in Appendix~C of \citetalias{pgs16}). To describe this process, we coin the term ``inter-plasmoid Compton scattering'' (IPCS).  Using results from simulations of relativistic reconnection and radiative transfer calculations, we demonstrate that the photons produced by a large plasmoid can increase the observed Compton dominance and high-energy emission from neighboring plasmoids by a factor of few. Our model predicts larger Compton ratios in several flares from BL-Lac like sources, which fall comfortably within the observed range, without requiring particle-dominated emission regions.

\section{Reconnection-Driven Plasmoids}
\label{sec:large_plasmoid_energy_density}

A plasmoid chain is composed of a large number of plasmoids with self-similar properties. A large majority of these plasmoids have sizes of $w_\perp / L \lesssim 0.05$, where $w_\perp$ is the plasmoid's transverse size (perpendicular to the reconnection layer and their direction of motion) and $L$ is the half-length of the reconnection layer, while only a few plasmoids obtain large sizes of $w_\perp / L \sim 0.1$ \citep{Uzdensky2010}. Large plasmoids move slowly (with non-relativistic speeds) in the reconnection layer,  grow from the coalescence with smaller plasmoids and direct accretion from the current sheet, and are long-lived with lifetimes equal to several light-crossing times of the layer (i.e. several $L/c$), as shown in   Fig.~\ref{fig:sigma_10_tracks} (thick lines). In this figure, and in what follows, we adopt the results from a  2D PIC simulation of reconnection in pair plasmas with $\sigma = 10$ \citep{sgp16}.

\subsection{Relative Motion of Plasmoids}
\label{sec:relative_motion}
\begin{figure}
\centering
\includegraphics[height=0.45\textwidth]{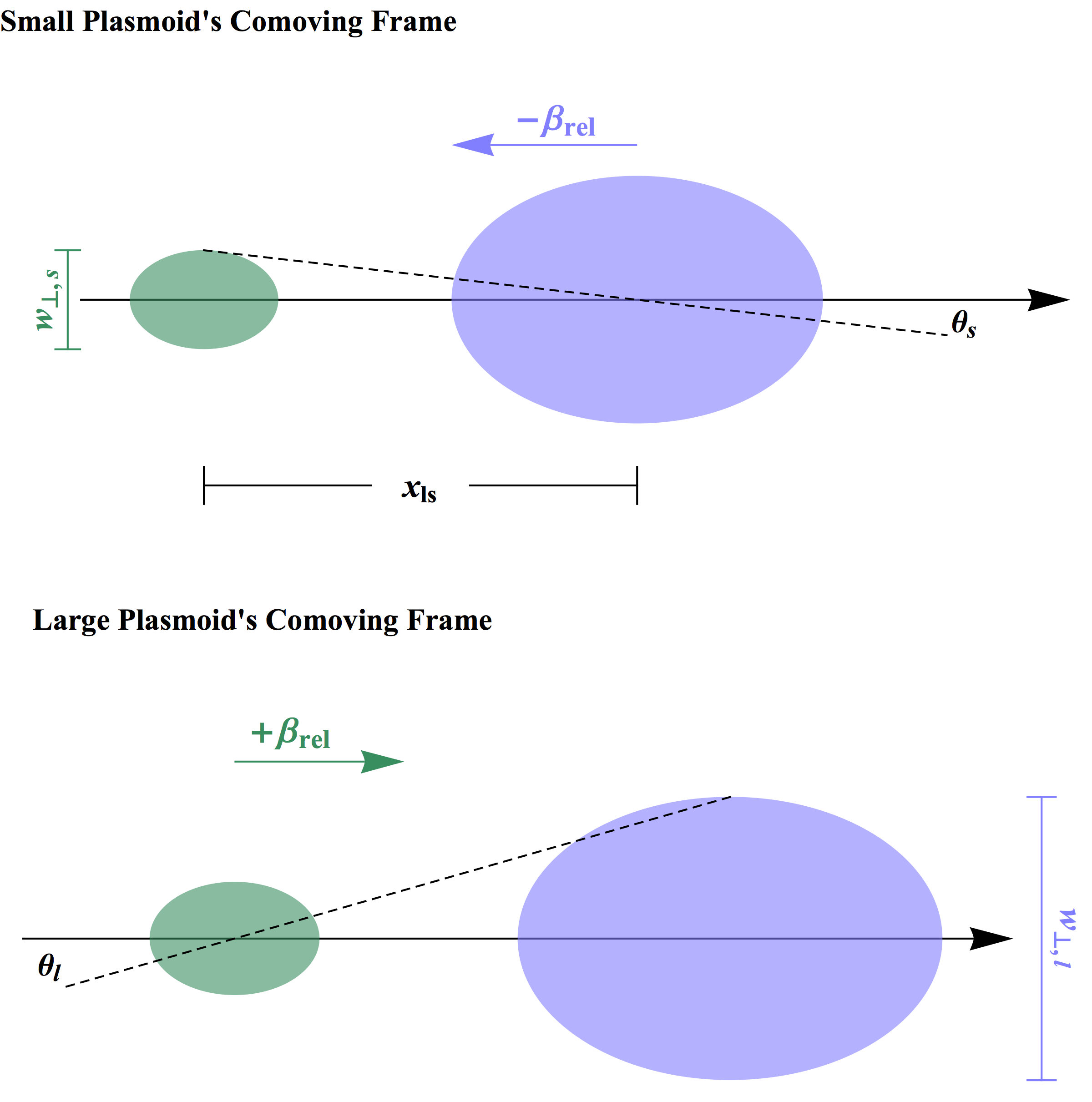}
\caption{Sketch illustrating the two reference frames used in the computation of the photon fields relevant for the IPCS process (see Sec.~\ref{sec:plasmoid_photon_energy_density}). The top panel displays the setup used to determine the photon energy of the small plasmoid as measured in the co-moving frame of the larger one (see eqn.~\ref{eqn:U_s_prime_prime}), while the bottom panel displays the geometry used to calculate the large plasmoid's energy density as measured in the smaller one's co-moving frame (see eqn.~\ref{eqn:U_l_prime}). In both panels, the black arrow represents the direction of bulk plasmoid motion (in the layer's frame), while the smaller arrows denote the direction of the relative velocity $\beta_{\rm rel}$ of one plasmoid in the co-moving frame of the other one. The angles marked on each panel are also measured in the respective reference frame. A coloured version of this plot is available online.}
\label{fig:sketch}
\end{figure}

Small plasmoids accelerate in the layer, reaching final speeds up to or exceeding\footnote{Small plasmoids which form in the vicinity of a large plasmoid can obtain dimensionless bulk four-velocities $\gtrsim \sqrt{\sigma}$ since the local Alfv{\'e}n speed is larger \citep{pcsg18}.} the Alfv{\'e}n velocity $v_{\rm A}/c= \sqrt{\sigma/(1+\sigma)}$. 
On the contrary, larger plasmoids move sub-relativistically with Lorentz factors $\Gamma_l \sim 1$. As a result, the relative motion of smaller plasmoids trailing a larger plasmoid can be relativistic with dimensionless velocity (i.e. normalized to the speed of light) $\beta_{\rm rel} = (\beta_s - \beta_l)/(1-\beta_s \beta_l)$ and Lorentz factor $\Gamma_{\rm rel} = \Gamma_s \Gamma_l (1 - \beta_s \beta_l) \gtrsim 1$, where the subscripts $s$ and $l$ refer to quantities of the small and large plasmoid, respectively.
This is illustrated in the right panel of Fig.~\ref{fig:sigma_10_tracks}, where the dimensionless four-velocity $\Gamma_{\rm pl} \, \beta_{\rm pl}$ for each plasmoid in which IPCS is relevant is displayed in color.

\subsection{Photon Energy Density of Plasmoids}
\label{sec:plasmoid_photon_energy_density}
In what follows, we demonstrate that large plasmoids can be an additional source of soft photons to be up-scattered by smaller, fast, neighboring plasmoids. 

Each plasmoid emits isotropically in its co-moving frame with luminosity $L_{ph}$ and photon energy density $U_{ph} \approx L_{ph} \, t_{\rm esc}/V$, where the photon escape timescale is $t_{\rm esc} = w_\perp/2c$ and $V \approx \pi w_\perp^3 /4$ is the plasmoid's co-moving volume\footnote{We assume that the plasmoid's shape in three dimensions is an ellipsoid with the third dimension being $w_\perp$ (\citetalias{christie2019}). However, all radiative calculations performed in Sec.~\ref{sec:results} assume spherical geometry.}, giving $U_{ph} \approx 2\, L_{ph} /\pi w_\perp^2 c$. 

Here, we compute the contribution of a plasmoid's photon field in the frame of its neighboring plasmoid. More specifically, we
i) determine the photon energy density of the small plasmoid as measured in the co-moving frame of the larger one and ii) compute the energy density of the larger one in the co-moving frame of the small one. In what follows, we denote all quantities measured  in the co-moving frame of the small (large) plasmoid with a single (double) prime. We also adopt the subscripts $s$ and $l$ to denote quantities of the small and large plasmoid, respectively. A sketch of two neighboring plasmoids, in two reference frames used in the calculations that follow, is displayed in Fig.~\ref{fig:sketch}.

To compute the energy density of the smaller plasmoid ($U_s^{\prime}$) as measured in the co-moving frame of the larger one ($U_{s}^{\prime \prime}$), we begin with the invariance of the quantity $U(\epsilon, \mu)/\epsilon^3$ \citep{rybicki_lightman,Dermer1994}:
\eqb 
\label{eqn:U_s_prime_prime}
U_s^{\prime\prime} &=&\int  \int d\mu^{\prime \prime} d\epsilon^{\prime \prime} U_s^{\prime \prime}(\epsilon^{\prime \prime}, \mu^{\prime \prime}) \\ \nonumber
& = &\Gamma^2_{\rm rel}\int \int d\mu^\prime d\epsilon^\prime U_s^\prime(\epsilon^\prime, \mu^\prime) (1+\beta_{\rm rel}  \mu^\prime)^2 \\ \nonumber 
& = & \Gamma_{\rm rel}^2 \, U^\prime_{s}\int_{\mu^\prime_s}^1 d\mu^\prime (1+\beta_{\rm rel}\mu^\prime)^2,
\eqe 
where we used the transformation of the angle and energy, namely $d\mu^{\prime \prime}= d\mu^\prime \Gamma_{\rm rel}^{-2} \left[1+\beta_{\rm rel} \mu^{\prime}\right]^{-2}$ and $\epsilon^{\prime \prime} = \Gamma_{\rm rel} \epsilon^\prime (1+\beta_{\rm rel} \mu^\prime)$. Here, $\mu^\prime \equiv \cos\theta^\prime$, where $\theta^\prime$ is the angle between the direction of  plasmoid motion and that of photons emitted by the small plasmoid (see top panel in Fig.~\ref{fig:sketch}), $\mu_{s}^\prime = \cos \theta_s^\prime \equiv x_{ls} /\sqrt{x_{ls}^2 + (w_{\perp,s} /2)^2}$, and $x_{ls}$ is the separation distance of the plasmoid centers. 
We chose to perform the integration in the small plasmoid's rest frame where the photon field is assumed to be isotropic, i.e. $U^\prime_s(\mu^\prime, \epsilon^\prime)= U^\prime_s(\epsilon^\prime)/2$ \citep{Dermer1994}. An additional factor of $2$, which cancels the factor $1/2$ from the isotropy of the radiation field, comes from the integration over angle (i.e. from $-\theta^\prime_{s}$ to $\theta^\prime_s$), which considers the radiation from both sides of the plasmoid (see Fig.~\ref{fig:sketch} for reference). The integral in eqn.~\ref{eqn:U_s_prime_prime} results in $U_s^{\prime \prime} = f_s^\prime \, U_s^\prime$, where \eqb
\label{eqn:f_s}
f_{s}^\prime \equiv \Gamma_{\rm rel}^2 \, [1 - \mu_{s}^\prime + \beta_{\rm rel} (1 - \mu_{s}^{\prime 2}) + \beta_{\rm rel}^2 (1 - \mu_{s}^{\prime 3})/3].
\eqe
In the limit of no relative motion (i.e. $\beta_{\rm rel}\rightarrow 0$ and $\Gamma_{\rm rel}\rightarrow 1$) and of large separation distance (i.e. $x_{ls} \gg w_{\perp,s}$), the expression above simplifies to the expression for the energy density of a point source at distance $x_{ls}$, namely $U^{\prime \prime}_s \approx L_s/ 4\pi x_{ls}^2 c$. Similarly, the energy density of the large plasmoid ($U_l^{\prime \prime}$) as measured in the co-moving frame of the smaller one ($U_{l}^\prime$) can be determined as  
\eqb 
\label{eqn:U_l_prime}
U_l^{\prime} &=&\int  \int d\mu^{\prime} d\epsilon^{\prime} U_l^{\prime}(\epsilon^{\prime}, \mu^{\prime}) \\ \nonumber
& = &\Gamma^2_{\rm rel}\int \int d\mu^{\prime \prime} d\epsilon^{\prime \prime} U_l^{\prime \prime}(\epsilon^{\prime \prime}, \mu^{\prime \prime}) (1 - \beta_{\rm rel}  \mu^{\prime \prime})^2 \\ \nonumber 
& = & \Gamma_{\rm rel}^2 \, U^{\prime \prime}_{l}\int_{-1}^{-\mu_l^{\prime \prime}} d\mu^{\prime \prime} (1 - \beta_{\rm rel}\mu^{\prime \prime})^2,
\eqe
where $\mu_{l}^{\prime \prime} = \cos\theta_l^{\prime \prime} \equiv x_{ls} /\sqrt{x_{ls}^2 + (w_{\perp,l} /2)^2}$ (see bottom panel in Fig. ~\ref{fig:sketch}). The integral in eqn.~\ref{eqn:U_l_prime} then leads to $U_l^\prime = f_l^{\prime \prime} \, U_l^{\prime \prime}$, where 
\eqb
\label{eqn:f_l}
f_{l}^{\prime \prime} \equiv \Gamma_{\rm rel}^2 \, [1 - \mu_{l}^{\prime \prime} + \beta_{\rm rel} (1 - \mu_{l}^{\prime \prime 2}) + \beta_{\rm rel}^2 (1 - \mu_{l}^{\prime \prime 3})/3].
\eqe

\begin{figure}
\centering
\includegraphics[height=0.55\textwidth]{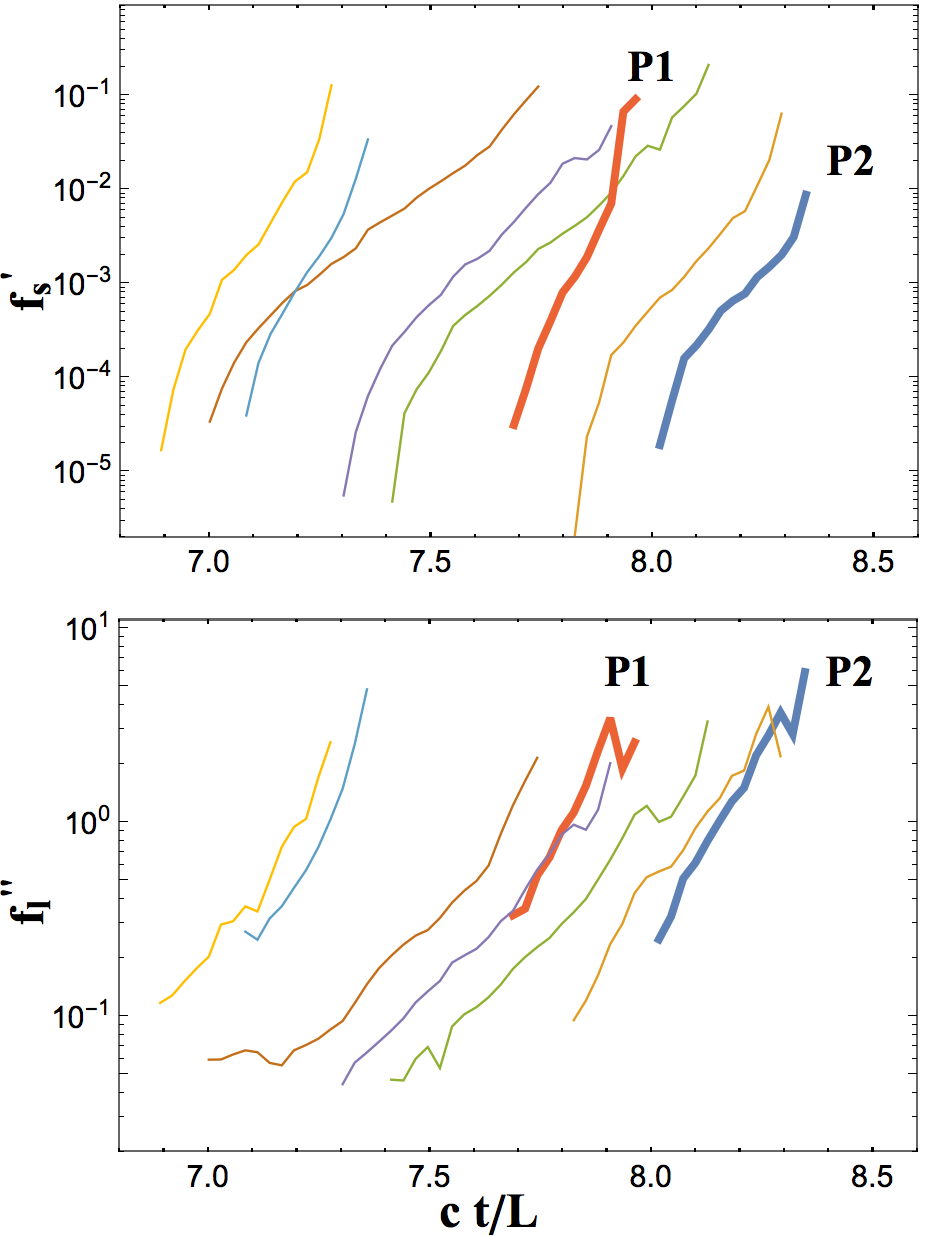}
\caption{Temporal evolution of the geometrical quantities $f_s^\prime$ (top panel, see eqn.~\ref{eqn:f_s}) and $f_l^{\prime \prime}$ (bottom panel, see eqn.~\ref{eqn:f_l}) for a subset of small plasmoids which trail behind a larger one (see right panel of Fig.~\ref{fig:sigma_10_tracks}). Because $f_l^{\prime \prime} \gg f_s^{\prime}$ for the plasmoids considered here, we can conclude that IPCS is relevant only for the smaller plasmoids.
The thick lines denote those plasmoids (their tracks denoted as P1 and P2 in Fig.~\ref{fig:sigma_10_tracks}) for which we later compute  spectra and light curves including the effects of IPCS (see Fig.~\ref{fig:small_plasmoids_SED_LC}). A coloured version of this plot is available online.}
\label{fig:fs_fl_plot}
\end{figure}
The two geometrical quantities $f_s^\prime$ and $f_l^{\prime \prime}$ are displayed in Fig.~\ref{fig:fs_fl_plot} for a small subset of plasmoids within the reconnection layer (see right panel of Fig.~\ref{fig:sigma_10_tracks} for reference). Both quantities show a similar temporal trend in which they gradually increase with time. This stems from the fact that $\Gamma_s$, and thereby $\Gamma_{\rm rel}$, continually increases with time until the small plasmoid merges with the large plasmoid lying ahead of it. The large difference in the values of $f_s^\prime$ and $f_l^{\prime \prime}$, spanning several orders of magnitude, originates from the large difference between the two plasmoid sizes, with respect to the distance between their centers $x_{ls}$. Here, the large plasmoid has size $w_{\perp, l} \sim 0.3 \, L$ while the small plasmoids typically have $w_{\perp, s} \lesssim 0.05 \, L$. Moreover, the distance between the plasmoid centers is $x_{ls} \sim 0.1 \, L \gg w_{\perp, s}$ (see right panel of Fig.~\ref{fig:sigma_10_tracks}). This leads to $\mu_s^{\prime} \sim 1$ for the majority of the small plasmoid's lifetime, whereas $\mu_l^{\prime \prime}$ quickly deviates from unity as $w_{\perp , l}$ grows larger than $x_{ls}$.  Only when the two plasmoids sizes become similar, $f_s^\prime$ and $f_l^{\prime \prime}$ will become roughly equal.

\begin{figure}
\centering
\includegraphics[height=0.32\textwidth]{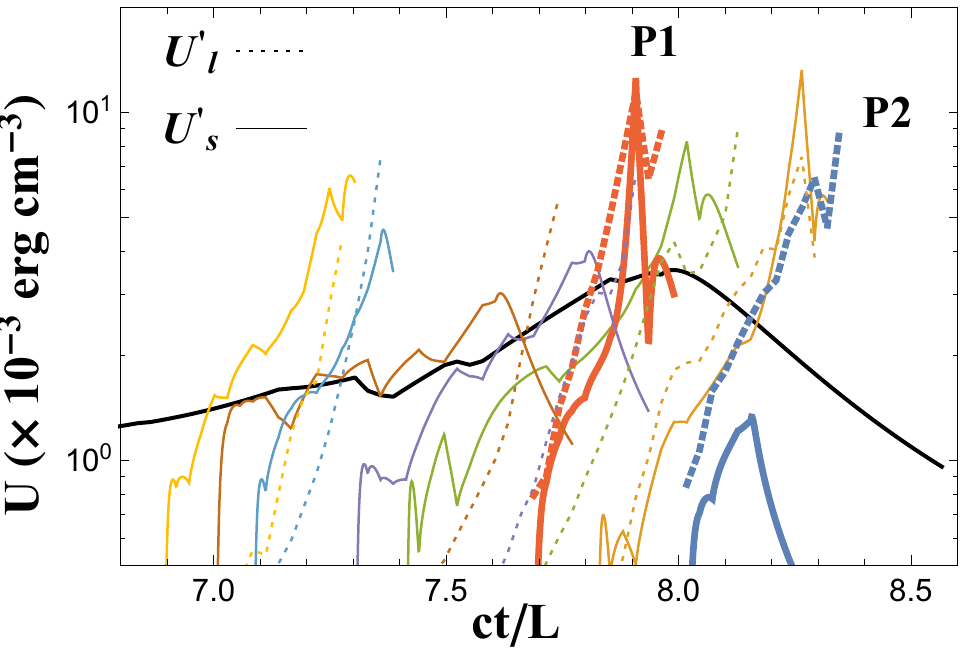}
\caption{Temporal evolution of synchrotron photon energy densities for the same subset of small plasmoids as in Fig.~\ref{fig:fs_fl_plot}. The photon energy density of the large plasmoid, as measured in the co-moving frame of the smaller ones, $U_l^\prime$ (colored dashed lines and eqn.~\ref{eqn:U_l_prime}), can, for some plasmoids, be larger than the photon energy density of the synchrotron photons in small plasmoids $U_s^\prime$ as measured in their co-moving frames (colored solid lines). Therefore, for small plasmoids whose value of $f_l^{\prime \prime} > 1$ (see bottom panel of Fig.~\ref{fig:fs_fl_plot}), IPCS can be a non-negligible process which can affect both the particle cooling and high-energy emission of smaller plasmoids. The thick lines denote those plasmoids (their tracks denoted as P1 and P2 in Figs.~\ref{fig:sigma_10_tracks} and~\ref{fig:fs_fl_plot})
for which we later compute  spectra and light curves including the effects of IPCS (see Fig.~\ref{fig:small_plasmoids_SED_LC}). For reference, we also show the energy density of the large plasmoid, as measured in its co-moving frame, $U_l^{\prime \prime}$ (solid black line). A coloured version of this plot is available online.}
\label{fig:f_tilde}
\end{figure}
For any small plasmoid that trails a larger one,
there is a competition between the synchrotron photons produced by the electrons in that plasmoid and the non-thermal radiation from neighboring larger plasmoids for the dominant source of seed photons for Compton scattering.
Here, and in what follows, we consider only the synchrotron radiation of the large plasmoid as an additional source of seed photons for Compton scattering by particles residing in  neighboring plasmoids. Higher energy photons produced in the large plasmoid through synchrotron self-Compton (SSC) would be up-scattered by particles in nearby plasmoids deep in the Klein-Nishina regime, where the Compton emissivity is suppressed.

\begin{figure*}
\centering
\includegraphics[height=0.528\textwidth]{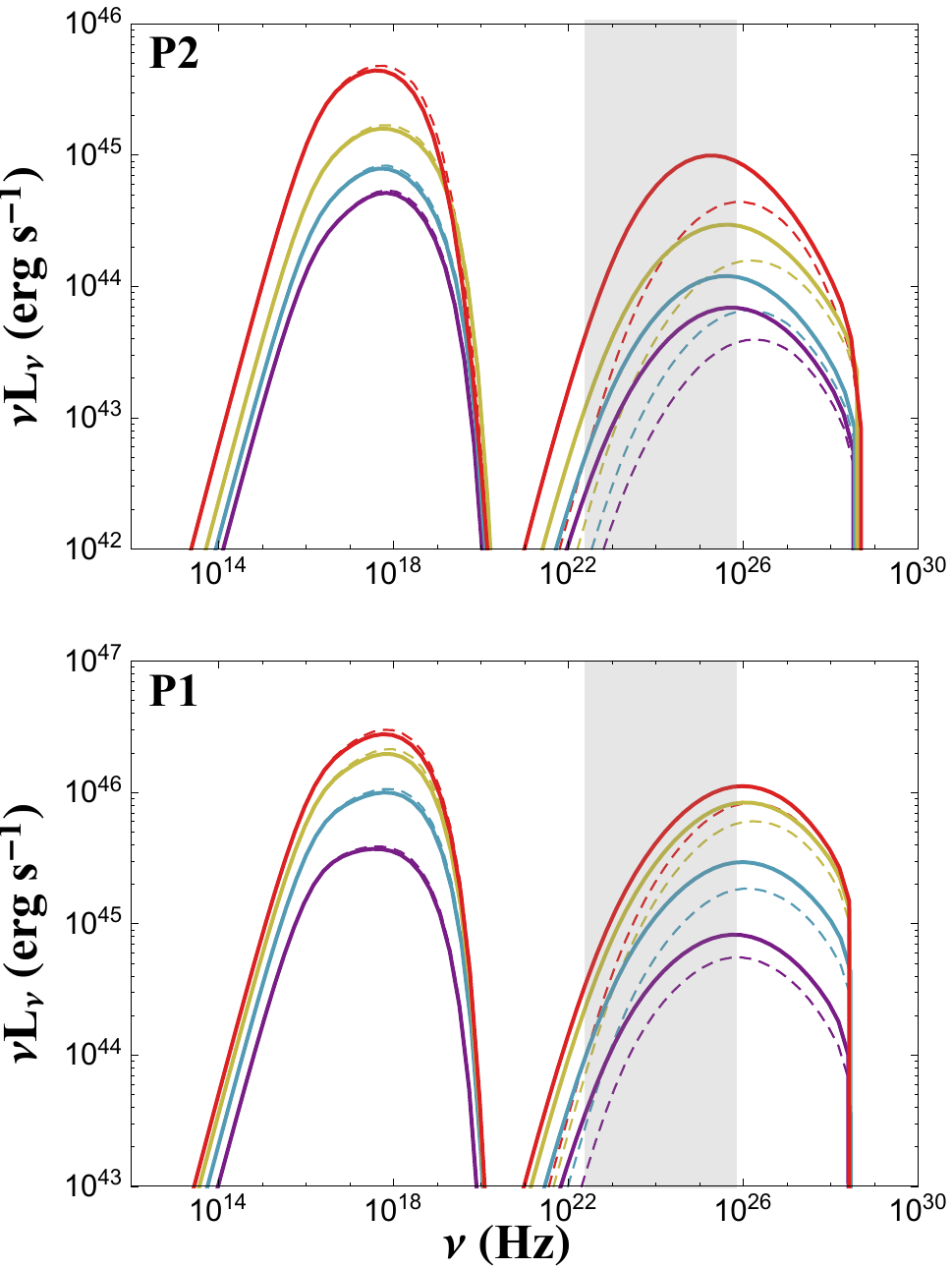}
\hspace{0.5 cm}
\includegraphics[height=0.52\textwidth]{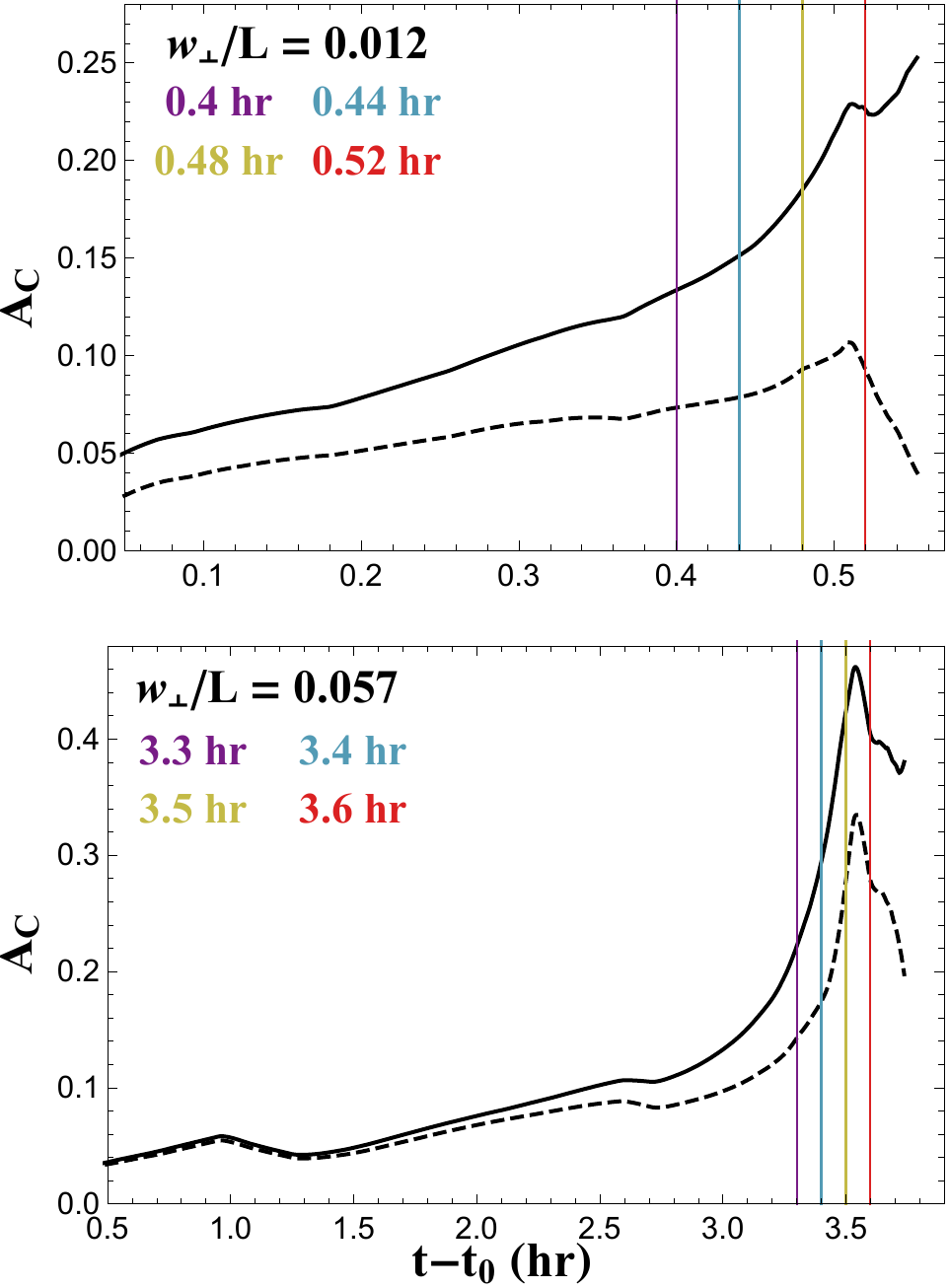}
\caption{Spectral energy distributions (SEDs; left column) and Compton ratios $A_C$ (right column) as seen by an observer, for two representative plasmoids (denoted by P1 and P2 in Figs.~\ref{fig:sigma_10_tracks},~\ref{fig:fs_fl_plot}, and~\ref{fig:f_tilde}). The final plasmoid sizes (normalized to $L$) are displayed in each panel of the right column while in all panels, the solid (dashed) lines denote the emission with (without) the inclusion of the non-thermal photons from the large neighboring plasmoid. The snapshots of the SEDs are color coded according to the time since the plasmoid birth, as measured by an observer aligned with the jet (see vertical lines in the right panels). 
The inclusion of the non-thermal photons from the large plasmoid results in an increase of the observed $\gamma$-ray luminosity (see gray band in left column for $0.1-300$~GeV \textit{Fermi}-LAT band), a broadening of the high-energy component, and an increase of the Compton ratio by a factor of $\sim 1.5-3$. A coloured version of this plot is available online.}
\label{fig:small_plasmoids_SED_LC}
\end{figure*}

Fig.~\ref{fig:f_tilde} displays $U_s^\prime$ and $U_l^\prime$ (denoted as the solid and dashed colored lines, respectively) for a subset of plasmoids (see right panel of Fig.~\ref{fig:sigma_10_tracks} and Fig.~\ref{fig:fs_fl_plot} for reference).
For one plasmoid in Fig.~\ref{fig:f_tilde} (e.g. thick blue curves), we find that $U_l^\prime \gtrsim U_s^\prime$ for the entirety of the small plasmoid's lifetime. For this plasmoid, the IPCS process becomes relevant, as the large plasmoid provides an important source of soft photons to be upscattered within the smaller plasmoid. 
For other plasmoids displayed in Fig.~\ref{fig:f_tilde} (e.g. thin yellow and green curves), $U_l^\prime \lesssim U_s^\prime$, since $f_l^{\prime \prime} < 1$ for the majority of a plasmoid's lifetime (see bottom panel of Fig.~\ref{fig:fs_fl_plot}). In these cases, IPCS can be neglected and these small plasmoids may well be considered as isolated objects. Lastly, there are several plasmoids (e.g. thin orange curves and thick red curves in Fig.~\ref{fig:f_tilde}) with $U_l^\prime \lesssim U_s^\prime$ at early times of their lifetime, but with $U_l^\prime \gtrsim U_s^\prime$ at later times.  This transition in the importance of the IPCS process will directly imprint itself on the radiative signatures of small plasmoids (for P1, see also  Fig.~\ref{fig:small_plasmoids_SED_LC}).

\section{Results}
\label{sec:results}

As discussed in the previous section, synchrotron photons from a large plasmoid can be an additional source of particle cooling for smaller trailing plasmoids. This can be imprinted on the observed spectra of smaller plasmoids. To study the effects of the time-dependent  photon fields from large plasmoids on the particle cooling and emission from smaller plasmoids, we perform numerical calculations using the radiative code described in \citetalias{christie2019}, which includes all the relevant physical processes.

\subsection{Inter-Plasmoid Compton Scattering}
\label{sec:numerical_analysis_monster_plasmoid}

We compute the emission from two representative plasmoids that form in the layer and move behind a larger plasmoid  (see lines marked with an asterisk in the right panel of  Fig.~\ref{fig:sigma_10_tracks} and with thick lines in Figs.~\ref{fig:fs_fl_plot} and~\ref{fig:f_tilde}). We use the same parameters as in the BL10 model presented in \citetalias{christie2019} (for model parameters, see fourth row in Table~1 therein), except for the magnetic field strength within plasmoids, which is now
$B \sim 0.6$~G.  Our results on the broadband photon emission from each plasmoid are presented in Fig.~\ref{fig:small_plasmoids_SED_LC}, with the left column displaying the temporal evolution of the SEDs and the right column displaying the Compton ratios\footnote{Here, $A_C$ is determined as the ratio of the peak luminosities of the high-energy to low-energy components (i.e. Compton and synchrotron, respectively) in a plasmoid's SED.} $A_C$. Solid and dashed lines denote the results with and without IPCS, respectively. A few things that are worth commenting on Fig.~\ref{fig:small_plasmoids_SED_LC} follow:
\begin{enumerate}
    \item The inclusion of the photons from the  large plasmoid  results in a higher Compton dominance $A_C$ (see right panel). For the two representative plasmoids, we find that $A_C\sim 0.3 - 0.5$ within their lifetimes. The largest $A_C$ value during the lifetime of a plasmoid is obtained around the time when its luminosity peaks, namely at the peak time of a flare. This time is also associated with roughly the time at which a plasmoid's Lorentz factor reaches its peak before it merges (see also Fig.~\ref{fig:fs_fl_plot}). Without IPCS, many small plasmoids, such as the one shown in the top row of Fig.~\ref{fig:small_plasmoids_SED_LC}, can only reach  low peak Compton ratios, e.g., $A_C \sim 0.1$ (see also \citetalias{pgs16} and \citetalias{christie2019}). Such low $A_C$ values
    are the combined result of the assumed equipartition between particles and magnetic fields (as dictated by PIC simulations of reconnection) and of the slow cooling\footnote{This refers to the regime where  the cooling timescale of electrons (with typical energy) is longer than the dynamical timescale of the plasmoid.} of electrons in, usually small, plasmoids\footnote{Particles within  small plasmoids typically reside in the slow-cooling regime, resulting in a low (i.e. $\sim 0.1$) peak Compton ratios (\citetalias{christie2019}).}.
    The IPCS process is important for plasmoids in which $U^\prime_l \gg U^\prime_s$ (see e.g., P2 in Fig.~\ref{fig:f_tilde}). Due to the  additional source of photons provided by the large neighboring plasmoid, the Compton ratio of the smaller plasmoid can increase by a factor of $\sim 3$, as shown in the top panel of Fig.~\ref{fig:small_plasmoids_SED_LC}. 
    However, IPCS can be less important for other plasmoids in the layer, as exemplified in the bottom row of Fig.~\ref{fig:small_plasmoids_SED_LC}. In this case, the presence of the larger plasmoid does not affect the high-energy emission of the trailing smaller plasmoid until late times within its lifetime (i.e., $t \gtrsim 2$~hr), when $U^\prime_l \gtrsim U_s^\prime$ (see see times $c \, t/L < 7.75$ for the thick red lines in Fig.~\ref{fig:f_tilde}). Inclusion of the IPCS process results in an increase of the Compton ratio by a factor of $\sim 1.5-2$ at and after the peak time of the flare (see lower right panel in Fig.~\ref{fig:small_plasmoids_SED_LC}).
    
    \item The inclusion of the synchrotron radiation from the large plasmoid as a seed for Compton scattering leads not only to an effective increase of the high-energy flux by a factor of $\sim 2 - 4$ (see \textit{Fermi}-LAT band, denoted as the gray band in Fig.~\ref{fig:small_plasmoids_SED_LC}), but also to a broadening of the high-energy spectrum (see left column of Fig.~\ref{fig:small_plasmoids_SED_LC}).
    The former is due to an increase in the available photon energy density while the latter is caused by the fact that the larger plasmoid, whose photons serve as seeds for IPCS, has a broader synchrotron spectrum than the smaller plasmoids (i.e. the large plasmoid has a lower minimum Lorentz factor in its  particle distribution, see eqn.~A4 in \citetalias{christie2019}). 
\end{enumerate}

\subsection{Application to BL Lac Sources}
\label{sec:blazar_application}
As discussed in Sec.~\ref{sec:numerical_analysis_monster_plasmoid}, the IPCS process can naturally increase the Compton ratio and luminosity of the high-energy component within the SEDs of fast moving plasmoids. An application of IPCS that is of particular importance is that of emission models of BL Lac sources. Typical BL Lacs have Compton ratios between $\sim 0.2 - 2$ \citep[see distributions provided in][]{fincke2013,Nalewajko2017}. When adopting single-zone emission models under the assumption of equipartition, as is expected for reconnection driven blazar flares, BL Lac-like models are found to have Compton ratios of $A_C \lesssim 0.1$. More recently, the BL Lac plasmoid-emission models of \citetalias{pgs16}, \citetalias{christie2019}, and \cite{Morris2019}, showed that the average $A_C$ is $\sim 0.2$, which falls near the lower bound of what is typically observed. 

A first step solution in alleviating this contradiction is including the IPCS process within the radiative calculations of small fast-moving plasmoids. By including this naturally occurring process, there can be a general increase in the Compton ratios of many plasmoids to values which fall within the general continuum of observations. However, some plasmoids (e.g. yellow curve in Fig.~\ref{fig:f_tilde}) which trail behind, and eventually merge with, a larger one do not show any significant changes within their emission when including the IPCS process. For these small plasmoids, their own synchrotron phootn energy density is greater than that of the large plasmoid's photon energy density. 
As such, their Compton ratios will remain unaffected by IPCS and these plasmoids can largely be considered as isolated objects within the relativistic blazar jet, an assumption taken in all previous plasmoid emission models. As all plasmoids in Fig.~\ref{fig:f_tilde} will produce short duration, luminous flares, one could expect variations in the observed Compton ratios between different flaring states. This is a common feature observed in many blazar sources when transitioning between different flaring states \citep[see][for variations in Mrk 421]{Acciari2011}. This can be a combination of the radiative efficiency of particle cooling via synchrotron emission and the relevance of IPCS through the plasmoid's relative motion with respect to large plasmoids. 

\section{Discussion and conclusions}
\label{sec:discussion}
Equipartition between magnetic fields and the radiating particles within plasmoids is a fundamental result of plasmoid-dominated relativistic reconnection. 
When invoking reconnection, and with it equipartition, very small Compton ratios are found for BL Lac sources (i.e. $A_C \lesssim 0.1$), in apparent contradiction with observations. We have shown that the non-thermal emission from a large plasmoid 
can serve as an additional seed photon field for Compton scattering by particles residing in its neighboring, smaller plasmoids which eventually merge with it. This inter-plasmoid Compton scattering (IPCS) process can lead to an increase of the high-energy flux emitted by smaller plasmoids and to Compton ratios as is typically observed in BL Lac sources (see Fig.~\ref{fig:small_plasmoids_SED_LC} for two representative plasmoids).

Here, we have mostly focused on a small subset of plasmoids trailing behind the largest plasmoid formed in the layer.
However, within a single reconnection event, there can be numerous sets of plasmoids  (see left panel of Fig.~\ref{fig:sigma_10_tracks}), each consisting of a single large plasmoid with $\sim 10 - 30$ smaller plasmoids trailing and eventually merging with it. Thus, the IPCS process may be relevant for a large fraction of the small plasmoids formed in the layer.
To obtain better statistics of IPCS importance within the layer, we can compare for each of these subsets the ratios $U_{s}^{\prime \prime}/U_l^{\prime \prime}$ and $U_{l}^{\prime}/U_s^\prime$ (see Sec.~\ref{sec:plasmoid_photon_energy_density} for derivation).
If particles within plasmoids are cooling efficiently through synchrotron radiation,
then $U_{ph} \simeq U_{\rm syn} \sim  U_e$, where $U_e$ is the energy density of injected relativistic electrons (\citetalias{pgs16}, see also eqn.~6 in \citetalias{christie2019}). 
The latter is found to be almost time-independent and approximately constant among plasmoids, regardless of their  sizes \citep[see also Fig.~5 of][]{sgp16}. Thus, the photon energy densities of the small and large plasmoids  are similar in their respective co-moving frames ($U_s^\prime \sim U_l^{\prime \prime}$, see solid lines in Fig.~\ref{fig:f_tilde} for reference) and the comparison of the ratios  $U_{s}^{\prime \prime}/U_l^{\prime \prime}$ and $U_{l}^{\prime}/U_s^\prime$ is mapped to a comparison of the parameters $f_{s}^\prime$ and $f_{l}^{\prime \prime}$. For those plasmoids which have sufficiently large $f_l^{\prime \prime}$ (i.e. values greater than a factor of a few), one could expect that IPCS is important. For example, of the eight small plasmoids presented in Figs.~\ref{fig:fs_fl_plot} and~\ref{fig:f_tilde}, half of them have large values of $f_l^{\prime \prime}$, and therefore $U_l^\prime \gtrsim U_s^\prime$ at some point within their lifetimes. By applying this same argument to $100$ plasmoids of those shown in the left panel of Fig.~\ref{fig:sigma_10_tracks}, we find that roughly half have sufficiently large values of $f_l^{\prime \prime}$ (i.e. $\sim 3 - 10$) and therefore $U_l^\prime \gtrsim U_s^\prime$. This signifies that for half of these small plasmoids, the IPCS process is an important mechanism for the high-energy emission, increasing the peak Compton ratios by a factor of $\sim 1.5 - 3$, such that their values fall within the observed distribution. 

Here, we discussed the role of IPCS within plasmoid emission models of BL Lac sources. Although plasmoids are also invoked to explain flaring episodes from other blazar subclasses, such as flat spectrum radio quasars (FSRQs, see \citetalias{christie2019} for a more detailed application), IPCS can be considered as a negligible process for small plasmoids. Emission models of FSRQs often adopt photon sources external to the jet (e.g. broad line region) to explain the observed large Compton ratios (typically, $A_C\gtrsim10$). However, the photon energy densities of these external fields are much greater that those produced even by the largest plasmoids in the layer. Thus, plasmoids in FSRQ emission models can be well considered as isolated objects within the jet and the radiative interactions between plasmoids can be ignored.

We conclude by stating a few simplifying assumptions that entered our calculations.
First, we neglect the fact that the photons from the large plasmoid are expected to induce Compton drag on the smaller plasmoids. As shown in \citet{beloborodov2017}, this effect is of secondary importance with respect to the cooling of electrons within small plasmoids. In other words, the removal of the plasmoid internal energy due to Compton cooling is expected to happen faster than the removal of its bulk energy due to Compton dragging. 
Second, we have neglected effects related to the inhomogeneous structure of plasmoids in regard to the plasmoid properties (e.g. magnetic field and particle density) as well as to the properties of the illuminating photon field (i.e. 
its intensity and spectrum do not depend on the location within the plasmoid).
We also consider that the plasmoid is moving as a rigid body even during mergers, which is likely a rough approximation.
Lastly, we neglect the finite light propagation time between the large, illuminating plasmoid and the smaller ones. This correction is of minor importance since the intrinsic properties of the large plasmoid (including its  overall emission spectrum and intensity) do not change much as the smaller plasmoids approach it.

When modeling blazar emission, the standard assumption for explaining the high-energy component of the spectrum is to invoke Compton processes seeded by synchrotron photons in the same emission region (SSC) or by photons external to the jet (external Compton). In the magnetic reconnection scenario for blazar emission, we have demonstrated the potential of another process that can naturally occur throughout the reconnection layer, namely the inter-plasmoid Compton scattering (IPCS). This process relies on the synchrotron photons produced by large plasmoids serving as seed photons for Compton scattering by particles in neighboring, smaller plasmoids. The inclusion of this effect can naturally lead to an increase in the Compton ratios during flares from BL-Lac like sources without requiring particle-dominated emission regions.

\section*{Acknowledgements} 
The authors would like to thank the anonymous referee for their constructive report. IC, MP, and DG acknowledge support from the Fermi Guest Investigation grant 80NSSC18K1745. MP also acknowledges support from the Lyman Jr.~Spitzer Postdoctoral Fellowship. LS acknowledges support from DoE DE-SC0016542, NASA ATP NNX-17AG21G, and NSF ACI-1657507. DG acknowledges support from the NASA ATP NNX17AG21G, the NSF AST-1910451 and the NSF AST-1816136 grants.


\bibliographystyle{mnras} 
\bibliography{bib.bib} 


\label{lastpage}
\end{document}